# Minimum Data Requirements and Supplemental Angle Constraints for Protein Structure Prediction with REDCRAFT

E. Timko[1], P. Shealy[1], M. Bryson[1], and H. Valafar[1]

[1]Department of Computer Science and Engineering, University of South Carolina, Columbia, SC

**Abstract** - *One algorithm to predict protein structure is the residual dipolar coupling based residue assembly and filter tool (REDCRAFT). This algorithm exploits an exponential reduction of the search space of all possible structures to find a structure that best fits a set of experimental residual dipolar couplings. However, the minimum amount of data required to successfully determine a protein's structure using REDCRAFT has not been previously investigated. Here we explore the effect of reducing the amount of data used to fold proteins. Our goal is to reduce experimental data collection times while retaining the accuracy levels previously achieved with larger amounts of data. We also investigate incorporating a* priori *secondary structure information into REDCRAFT to improve its structure prediction ability.*

## 1 Introduction

Over the past few years, the utility of residual dipolar couplings (RDCs) has increased precipitously. An important application of this new data source is structure determination of proteins [1-5]. This explosion in the use of RDC data for structure determination of macromolecules is due to the distinct advantages of RDCs over the traditional approach of obtaining distance constraints from NOE data. In general, RDCs are more precise and easier to measure, concurrently provide structural and dynamic information, and exhibit a direct relationship to critical structural parameters such as backbone torsion angles. Given the alignment of an unknown protein, a single RDC datum can limit the orientation of its corresponding internuclear vector to within two symmetrical cones. It would be difficult to extend a similar claim to a single NOE constraint per residue.

Additionally, the number of NOEs required for an unambiguous recovery of a structure is dependent on the structural complexity of the protein, which is unknown *a priori* to structure determination. The lack of an understanding of the required amount of data has a direct impact in the financial and temporal cost and success of protein structure characterization using NOEs. In contrast, RDCs are well suited to a theoretical understanding of required data levels independent of the structure's complexity. Because RDCs can directly constrain related torsion angles, it is no surprise that 2-6 RDCs per residue should suffice for successful determination of a protein backbone, regardless of its structural complexity. In addition, RDCs are measured with respect to a global reference (the common order tensor, under the assumption of molecular rigidity). This constrains errors to have only a local effect, preventing errors in one section of the molecule from affecting other sections. In addition, discrepancies in the molecule's rigidity can at least be evaluated, if not completely characterized. Neither advantage exists for NOEs.

Although analysis of RDC constraints is conceptually more difficult than distance constraints, their algorithmic implementation is computationally friendlier. Since NOE interactions can be observed between any two atoms on the protein backbone, piecewise structure determination becomes impossible. Therefore, the structure of the entire protein needs to be tackled all at once (including the side chains). The energy landscape observed during structure determination of the entire protein typically contains many local minima, and the high dimensionality of the search space and complexity of the energy landscape mean that approaches like simulated annealing exhibit difficulty in finding the optimal structure. RDCs, and their related constraints, offer the ability to construct fragments of the protein backbone incrementally, through the addition of one amino acid at a time. This leads to a computationally friendly approach that allows direct investigation of the protein backbone first, then addition of side chains at a later stage.

Properly designed RDC analysis tools for structure determination can become impervious to the above mentioned shortcomings of NOE-based structure determination algorithms while exploiting all of the advantages of RDC data. Because of their unique properties, RDCs play an increasingly important role in NMR structure determination. However, structure determination primarily based on RDC data requires new programs that operate in fundamentally different ways from those that use NOE data. One such algorithm is REDCRAFT [6], an approach that has previously been shown successful with modest data requirements. Here we investigate reducing the number of RDCs per residue that are required to effectively compute a protein structure. Reducing the required number of RDCs results in a reduction of data collection time as well as financial cost. We also investigate the effect of incorporating *a priori* knowledge of the protein's secondary structure into REDCRAFT, with the goal of improving its structure computation abilities.

## 2 Background and Method

### 2.1 Residual Dipolar Couplings

A number of recent structure determination algorithms have used residual dipolar couplings (RDCs) as a data source [1-5] . RDCs have also been used in studies of carbohydrates [1; 7], nucleic acids [8; 9], and proteins [1; 5; 10-14]. Although residual dipolar couplings (RDCs) were observed as early as 1963 [15], there has been a recent resurgence in their use, due mostly to improvements in inducing weak alignment in proteins [2]. Upon the reintroduction of order to an isotropically tumbling molecule, RDCs can be acquired very easily. The RDC interaction between two atoms in space can be formulated as shown in Eq. (1).

$$D_{ij} = \frac{-\mu_0 \gamma_i \gamma_j h}{(2\pi r)^3} \left\langle \frac{3\cos^2(\theta_{ij}(t)) - 1}{2} \right\rangle \quad (1)$$

In this equation, $D_{ij}$ denotes the residual dipolar coupling in units of Hz between nuclei $i$ and $j$, $\gamma_i$ and $\gamma_j$ are nuclear gyromagnetic ratios, $r$ is the internuclear distance (assumed fixed for directly bonded atoms) and $\theta_{ij}$ is the time-dependent angle of the internuclear vector with respect to the external magnetic field. The angle brackets signify time averaging.

Algebraic manipulation of Eq. (1) can lead to an alternate RDC formulation in matrix form (Eq. (2)).

$$D \propto X^T \times S \times X = X^T \times R \times S' \times R^T \times X \quad (2)$$

$$R = R_z(\alpha) R_y(\beta) R_z(\gamma) \quad (3)$$

$S$ and $S'$ are $3 \times 3$, symmetric, traceless matrices related to one another through a Jacobi transformation, and $S$ is the Saupe order tensor matrix [15]. The transformation $R$ (Eq. (2), Eq. (3)) encapsulates three Eulerian rotations $\alpha$, $\beta$, $\gamma$ in the Cartesian space and relates $S$ to $S'$.

### 2.2 Structural Fitness Calculation

While it is nontrivial to generate a protein structure given a set of residual dipolar couplings, it is straightforward to determine how well a given structure fits a set of RDCs. An RDC $d$ is first formulated as shown in Eq. (4).

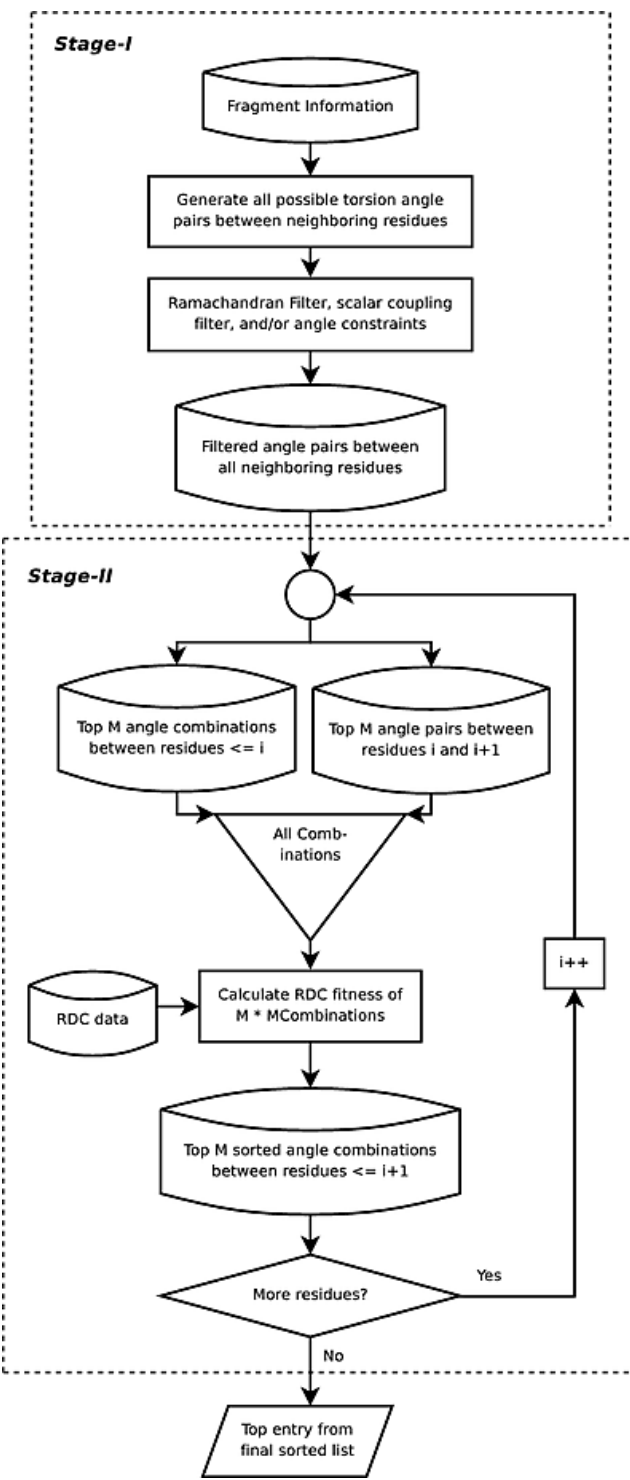

Figure 1: The REDCRAFT algorithm

$$d = x^2 S_{xx} + y^2 S_{yy} + z^2 S_{zz} + 2xy S_{xy} + 2xz S_{xz} + 2yz S_{yz} \quad (4)$$

$S_{xx}$, $S_{yy}$, $S_{zz}$, $S_{xy}$, $S_{xz}$, and $S_{yz}$ are elements of the $S$ matrix from Eq. (2). Given a set of $n$ RDCs for a protein, it is convenient to consider the matrix formulation given in Eq. (5), where $A$ is an $n \times 6$ matrix composed of atom coordinates and *D* is an *n x 1* vector of RDCs. Given this form, the pseudoinverse $A^+$ of *A* can be calculated using the singular value decomposition [16]. Eq. (6) then provides a fitness measure in units Hz of a protein structure to a set of RDCs, where a smaller fitness value indicates a better structure.

$$A S = D, \quad S = [S_{xx}\ S_{yy}\ S_{zz}\ S_{xy}\ S_{xz}\ S_{yz}]^T \quad (5)$$

$$Fitness = \sqrt{\sum_{i=1}^{n}(D_i - D'_i)^2}, \quad D' = A A^+ D \quad (6)$$

## 2.3 REDCRAFT

The REDCRAFT algorithm and its success in determining medium-resolution structures has been previously described in detail [6]. Here we present a brief overview. REDCRAFT's algorithm for calculating structures from RDCs constrains two separate stages. In the first stage (*Stage-I*), a list of all possible discretized torsion angles is created for every pair of adjoining peptide planes. Some angles in these lists are removed based on filters such as Ramachandran space [17] and scalar couplings [18]. The torsion angles are then ranked based on fitness to the RDC data. These lists of potential angle configurations are used to reduce the search space for the second stage.

*Stage-II* of the analysis starts with the first two peptide planes of the protein. Every possible combination of angles from *Stage-I* between peptide planes one and two are evaluated for fitness with respect to the collected data, and the best *M* candidate structures are selected, where *M* is the search depth. These *M* angle combinations are then combined with every possible set of angles connecting the next peptide plane to the protein. Each of these candidate structures is evaluated for fitness and the best *M* are again selected and carried forward. All angle combinations worse than the best *M* combinations are eliminated, thus removing an exponential number of candidate structures from the search space. This is repeated iteratively, incrementally adding peptide planes until the entire protein has been constructed. These two stages are illustrated in Figure 1.

The choice of search depth *M* has a large impact on the final structure determined with this algorithm. If *M* is too small, the global optimal angle configuration will likely be eliminated from the search space. However as *M* increases, the memory and runtime required also dramatically increases. Using a search depth of 2000 has provided acceptable structures using both computationally generated data and purely experimental data, with a runtime in the order of hours on a typical desktop computer, and is the value used for all experiments discussed here.

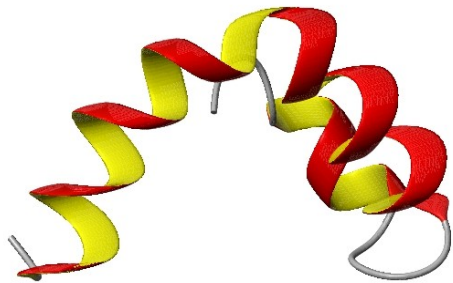

Figure 3: The structure of 1A1Z for the first 40 residues.

The number of RDCs required to correctly fold a protein with REDCRAFT has not been previously examined in a systematic manner. Here we investigate the effect of reducing the available RDCs on the quality of the resulting computational structure. Collecting fewer RDCs per peptide plane can substantially reduce data collection times. In particular, $^{15}$N-$^1$H RDCs are easily collected because they avoid expensive $^{13}$C labeling. Furthermore, $^{15}$N-$^1$H RDC values are typically large in magnitude, reducing the effect of measurement error. $C^\alpha$-$H^\alpha$ RDCs are large in magnitude but require $^{13}$C labeling, complicating sample preparation. RDCs for additional vectors can be collected, but with a decreasing utility and at a greater expense.

We also investigate the impact of incorporating secondary structure information into Stage-I in an attempt to improve structure prediction accuracy. Here we use secondary structure knowledge from the known structure of 1A1Z. In practice, one could obtain secondary structure information using, e.g., chemical shifts [19] or primary sequence [20].

## 3 Results and Discussion

We analyzed REDCRAFT's performance using the 83-residue α-helical domain with PDB code 1A1Z. For this work, experimental data is only computed for the first 40 residues of 1A1Z to reduce computation time. This portion of 1A1Z is shown in Figure 3.

|  | *Ram-II* | | *All space* | |
| --- | --- | --- | --- | --- |
| *Number of Vectors* | REDCRAFT RMSD score (Hz) | Backbone RMSD (Å) | REDCRAFT RMSD score (Hz) | Backbone RMSD (Å) |
| 1 | 0.305 | 7.424 | 0.324 | 9.656 |
| 2 | 0.977 | 3.238 | 0.834 | 3.177 |
| 3 | 0.780 | 3.209 | 0.803 | 3.448 |
| 6 | 0.708 | 3.390 | 0.719 | 3.375 |

Figure 2: RMSD scores for Ram-II and All Space, no secondary structure information.

| Number of Vectors | Ram-II | All Space |

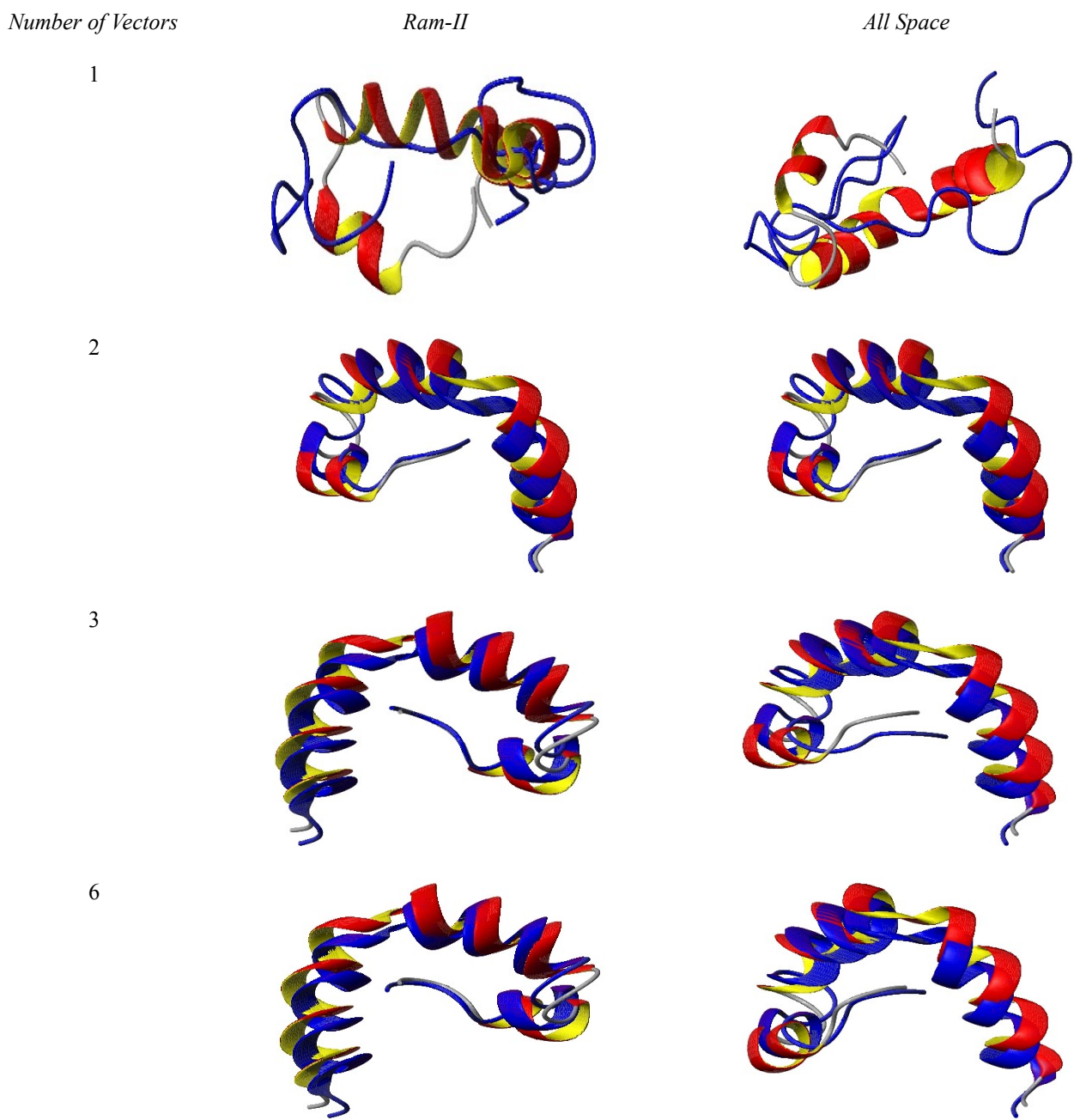

Figure 4: Overlays for structures computed using reduced amounts of data.

We define a computed structure's REDCRAFT RMSD score to be the RMSD between the experimental RDCs and the structure's back-calculated RDCs. This value is in units of Hertz. The backbone RMSD is defined as the RMSD of the distance between backbone atoms from the computational and true structures and is angstroms.

Stage-I of REDCRAFT requires a set of acceptable (φ, ψ) angle pairs for each peptide plane. Two strategies used previously are Ramachandran Level II (*Ram-II*), which restricts angles to plausible regions of the Ramachandran plot, and *All space*, which imposes no restrictions on allowed angles. Here we introduce a third strategy: selectively restricting angles of 1A1Z to the α-helical region of the Ramachandran space.

We analyze the impact of using data from different numbers of internuclear vectors. RDCs for each vector were computed in two independent alignment media; all results use data from both media. The vectors were chosen based on ease of data collection and the information content of the vector's RDCs. For peptide plane $i$, we use the following vectors: for one vector per peptide plane, we use N($i$)-H($i$) RDCs; for two vectors, N($i$)-H($i$) and C$^\alpha$($i$)-H$^\alpha$($i$); for three vectors, N($i$)-H($i$), C$^\alpha$($i$)-H$^\alpha$($i$), and C($i$)-N($i$); and for six vectors, N($i$)-H($i$), C$^\alpha$($i$)-H$^\alpha$($i$), C($i$)-N($i$), C($i$)-H($i$), H$^\alpha$($i$)-H($i$), and H$^\alpha$($i$-1)-H($i$). All RDCs are computed using REDCAT [21] and contain noise uniformly added in the range ± 1Hz.

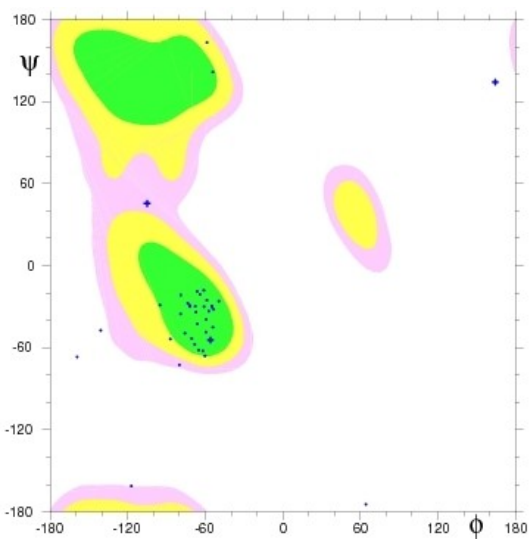

Figure 5: The Ramachandran Plot for the first 40 residues of 1A1Z.

## 3.1 Experimental Data without Supplemented Angle Constraints

Figure 2 shows RMSD scores for the experimental structure obtained using only one vector. The REDCRAFT RMSD score is lower than 0.4 Hz for both angle constraints, but the backbone RMSD is more than 7 Å for each. The computed structures are overlaid with the true structure in Figure 4. (For reference, the experimental structures created using all six vectors are shown in Figure 4, and their RMSD data can be found in Figure 2.)

The REDCRAFT RMSD is lower than the 1 Hz error for the RDCs, implying that the structures fit the data well and suggesting that high-quality structures were found. However, the overlays clearly show that the computed structure shows little resemblance to the true structure. In particular, the computed structure contains no α-helices, while the true structure has three. The structure determined using no angle constraints is worse than the structure created using Ram-II. This is due to a loss of true φ and ψ angles while running REDCRAFT. Because REDCRAFT only retains a fixed number of candidate structures at each iteration, the true angles may be lost at an intermediate step, resulting in a poor final structure. These results suggest that data from one vector is insufficient to determine protein structure.

Figure 2 shows RMSDs for structures obtained using RDCs from both N-H and C$^\alpha$-H$^\alpha$ vectors. The REDCRAFT RMSD score is below 1.0 Hz for both angle constraints, with a backbone RMSD is slightly more than 3.0 Å for each. Figure 4 shows the experimental and true structures using Ram-II and no angle restraints, respectively.

When using two vectors for REDCRAFT, the experimental structure is a much better match to the true structure. This is true with both types of angle constraints. The structure determined with no angle constraints is actually slightly better than that using Ram-II because some torsion angles reside outside of the Ramachandran space for the true 1A1Z structure (Figure 5). Although the REDCRAFT RMSD is higher than the RMSD for structures created using one vector, the quality of the structures produced is much better.

Figure 2 contains the REDCRAFT RMSD score for structures created using RDCs from three vectors. The REDCRAFT RMSD score is lower than 1.0 Hz, but the backbone RMSD is greater than 3.0 Å. The overlay figures are shown in Figure 4.

After incorporating the third vector, the structure is slightly worse than the structure created with two vectors, but there is still a good match between the experimental and true structures. However, at residue 29, the structure begins to diverge from the true structure. The REDCRAFT RMSD is still significant because it is less than the 1 Hz error set in the original two alignment media. C-N has a limited range of observable data values, so the additional information provided by this RDC is consumed by the 1 Hz error, resulting in a slightly worse structure.

We have presented two measures of structure fitness: REDCRAFT RMSD and backbone RMSD. When searching for the best structure, REDCRAFT uses only its RMSD score as a guide. However, because the true structure is also available, the backbone RMSD between the best computed structure and the true structure is more insightful into REDCRAFT's performance. These two measures of error may differ because when insufficient data is used in structure calculation, the search space is insufficiently constrained, resulting in a solution that fits the experimental data very well but is a poor representation of the true structure. Backbone RMSD provides better insight into the robustness of the computed structure.

| Number of Vectors | REDCRAFT RMSD score (Hz) | Backbone RMSD (Å) |
|---|---|---|
| 1 | 0.893 | 5.536 |
| 2 | 6.556 | 7.851 |
| 3 | 3.580 | 5.984 |
| 6 | 2.447 | 6.160 |

Figure 6: RMSD scores with supplemental secondary structure angle restraints.

## 3.2 Experimental Data with Supplemented Angle Constraints

Limited RDC data can be supplemented with angle restraints based on secondary structure information for the protein. 1A1Z consists solely of α-helices and turns between helices. For 1A1Z, most α-helices, φ angles fall in the range -75º ± 35º, and most ψ angles fall in the range -40º ± 40º. These ranges were used to restrict the search space for angles in α-helices. For turns, Ram-II space is used to restrict φ and ψ angles.

When using only N-H RDCs, as shown in Figure 6, the REDCRAFT RMSD score is less than 1 Hz, but the backbone RMSD is greater than 5 Å. As shown in the overlay (Figure 7), the structure created using only N-H vectors is only close to the first part of the true structure's first α-helix, after which the two diverge. A REDCRAFT score less than 1 Hz implies that the computed structure is of high quality. However, the computed structure begins to deteriorate after residue 16 when REDCRAFT attempts to compensate for the structure starting at that point. Several residues, starting at residue 16, lie far outside of the ideal α-helical angles, which may explain these results.

As seen in Figure 6, when using RDCs from both N-H and $C^α$-$H^α$ vectors, both the backbone RMSD and REDCRAFT RMSD score are high: greater than 6 Hz and 7 Å respectively. Figure 7 shows the computed structure overlaid with the true structure. Both the REDCRAFT RMSD score and the backbone RMSD suggest that the structure is not robust at all. The experimental structure in Figure 7 deviates more from the true structure than the computed structure using only one RDC. The experimental structure's secondary structure also does not match that of the true structure. It is likely that, upon reaching residue 16, REDCRAFT discarded the candidate structures that matched the data well in an attempt to compensate for the unusual angles encountered in that region, which were disallowed in Stage-I.

Figure 6 shows that the REDCRAFT RMSD has decreased a fair amount by adding a third vector, but it still remains above 3.0 Hz. The backbone RMSD has also decreased, but is still more than 5.0 Å. The overlays are shown in Figure 7. When adding this third vector (C-N, N-H, and $C^α$-$H^α$ total), the structure has recovered some of its robustness as shown in Figure 7. Both RMSD scores have decreased, suggesting that more data has helped REDCRAFT find a better structure compared to the structure that was determined using two vectors. Again, the computed structure follows fairly closely to the first α-helix of the true structure until it reaches residue 16. After residue 16, the computed structure does not follow the true structure.

| Number of Vectors | Overlay |
|---|---|
| 1 | |
| 2 | |
| 3 | |
| 6 | |

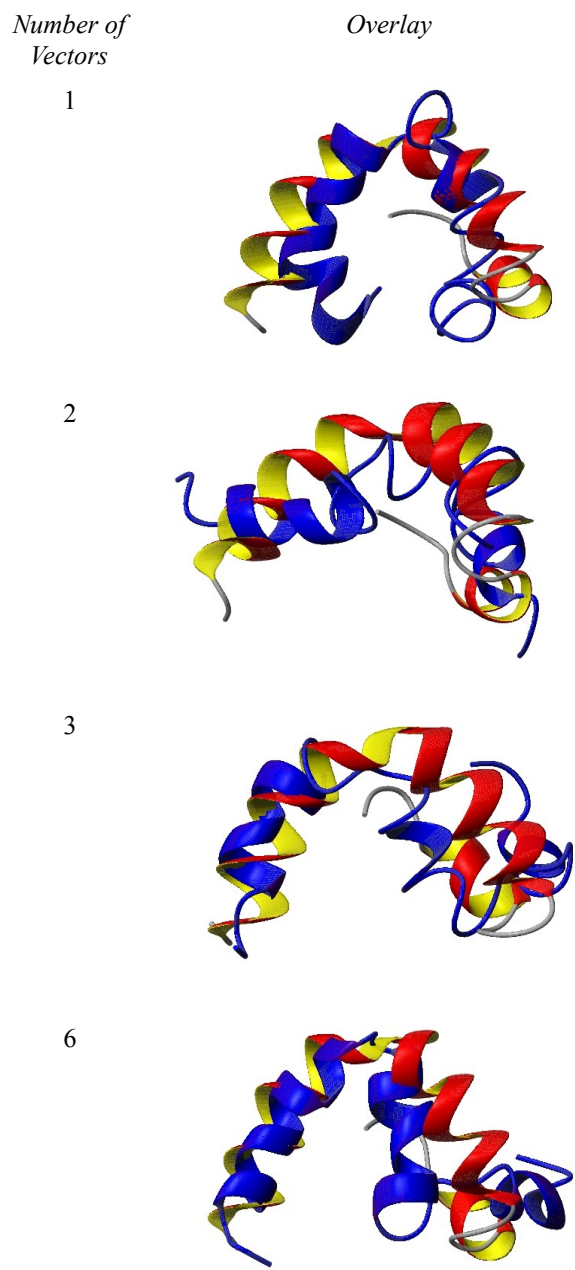

Figure 7: Overlay figures created using supplemental angle constraints.

Even when using all six vectors, shown in Figure 7 for reference, the same problem occurs with the structure only following closely to the first α-helix. The respective RMSDs are shown in Figure 6, and even though the REDCRAFT RMSD drops to just over 2 Hz, the backbone

RMSD rises slightly to just over 6 Å.

# 4 Conclusion

Overall, we have found that at least two vectors are necessary to get a robust structure using our methodology for protein structure determination. RDCs collected from one vector in two alignment media do not appear to be sufficient to determine the backbone structure of the protein 1A1Z, so other methods must be utilized in order to supplement that lack of data. Incorporating additional data is one successful strategy. However, after two vectors, no improvement occurred.

Another strategy we tried is to incorporate additional information in the form of secondary structure knowledge. However, using angle constraints related to the secondary structure was unsuccessful. Although the computed structure showed some correlation with the true structure, in general the computed structure was of poor quality. This is likely due to a few angles in α-helix regions falling outside the typical values for helices. However, secondary structure information is an important source of additional information, so we plan to investigate alternative methods for including it in the REDCRAFT algorithm.